\documentclass{article}
\usepackage{fleqn,ams}
\usepackage{amssymb}
\usepackage{verbatim}

\def\stackunder#1#2{\mathrel{\mathop{#2}\limits_{#1}}}
\newcommand{\Osymbol}[1]{\mbox{{\Large $\bigcirc$}}\hspace{-1.11em}{\mbox{\small  #1}}}
\newcommand{\Boxf}[2]{\parbox{#1cm}{\vskip 4pt {\footnotesize #2}\vskip
4pt\hfill}}

\newcommand{\Boxn}[2]{\parbox{#1cm}{\vskip 4pt  #2\vskip
4pt\hfill}}
\begin{document}
\begin{center}
{\bf\Large The Relativistic kinetics of gravitational waves
collisional
damping in hot Universe}\\[12pt]
Yu.G.Ignatyev, V.Yu.Shulikovsky\\
Kazan State Pedagogical Institute,\\[12pt] Mezhlauk str., 1, Kazan
420021, Russia
\end{center}

\begin{abstract}The article is a translation of authors paper
\cite{YuSh} printed earlier in the inaccessible edition and
summarizing the results of research of gravitational waves damping
problem in the cosmologic plasma due to the different interactions
of elementary particles.\end{abstract}

\section*{Introduction}The evolution of the cosmological gravitational waves (GW)
in the Universe, in case that perfect fluid or a collisionless gas
is considered as a material medium model, is researched
sufficiently \cite{Yukin1} - \cite{Yukin4}. Let us note basic
results:

\begin{itemize}
\item[-] perfect fluid doesn't change a vacuum character of the
propagation of the gravitational radiation;
\item[-] dispersion is essential only for super-density objects.
\end{itemize}
There were made an attempts to provide an analysis of evolution
for dissipative mediums (see, for example
\cite{Yukin5},\cite{Yukin6}). In authors works \cite{Yukin4},
\cite{Yukin7} --- \cite{Yukin9} the damping of gravitational
radiation during the propagation in the collisionless gas was
considered in the context of kinetic approach. Given paper may be
considered as a concluding in the whole cycle of researches.

\section{Self-consistent model of weak gravitational waves evolution
in the Friedmann Universe} In the spatially flat Friedmann world
with a metric:
$$d\stackrel{o}{s}\ \!\! ^2 =\
\stackrel{o}{g}_{ik}dx^idx^k=$$
\begin{equation}\label{1.1}
=a^2(\eta)[d\eta^2-(dx^1)^2 - (dx^2)^2 - (dx^3)^2]
\end{equation}
gravitational waves are described by metric perturbation:
\begin{equation}\label{1.2}
h_{ik} = g_{ik} -\stackrel{o}g_{ik} =
\tilde{h}_{ik}(\eta)e^{-in_{\alpha}x^{\alpha}}; \qquad
\alpha,\beta = \overline{1,3},
\end{equation}
where $n_{\alpha} = \rm{const}$ - wave vector GW\footnote{Here and
further the Greek letters run the values from 1 to 3; Latin - from
1 to 4.}; it's reference components, measuring by synchronic
observer in metric are (\ref{1.1}),
$$k_{(\alpha)} = \frac{n_{\alpha}}{a};\qquad k = \frac{n}{a},$$
where $n = \sqrt{n^2_1 + n^2_2 + n^2_3}.$ It is obvious that in
the flat Friedmann world waves of any length $\lambda$ can be
counted like short:
\begin{equation}\label{1.3}
\lambda \ll R \rightarrow \infty
\end{equation}
where $R$ ---the radius of background curvature. Then on the
$h_{ik}$ additional conditions \cite{Yukin1} of graduation and
tracelessity  can be imposed:
\begin{equation}\label{1.4}
h_{i4} = 0,\quad h = 0,\quad h^{ik}_{\ ,k} = 0\rightarrow
h_{\alpha}^\beta n_{\beta} = 0;
\end{equation}
here and further the indexes lift and lower by the background
metric $\stackrel{o}{g}_{ik}$, $h=g_{ik}h^{ik}$. How it will be
seen from the further, the perturbation of the momentum-energy
tensor (MET) of medium, determining by the GW appearance is
representable in the form:
\begin{equation}\label{1.5}
\delta T_{ik} = \tau(\eta)h_{ik}.$$
\end{equation}
Then the system of linearized by $h$ Einstein equations can be
written in the form
$$\psi'' + n^2\psi - 2\left(\frac{a'}{a}\right)^2\psi  -
\left(\frac{a''}{a}\right)\psi+ 16\pi \psi a^2\tau = 0,$$ where in
further we will set:
\begin{equation}\label{1.6}
\tilde{h}_{\alpha \beta}(\eta) = a(\eta) S_{\alpha
\beta}\psi(\eta),
\end{equation}
at that $S_{\alpha \beta} = \mbox{const}$. Taking into account the
background Einstein equations%
\begin{equation}\label{1.7}
\left(\frac{a'}{a}\right)^2 = \frac{8\pi \varepsilon a^2}{3},\,
\frac{a''}{a} = -\frac{4\pi}{3}a^2(\varepsilon + 3p) -
\frac{8\pi}{3}a^2\varepsilon ,
\end{equation}
finally we will receive the evolution of {\it GW conform amplitude
} equation $\psi(\eta)$:
\begin{equation}\label{1.8}
\psi'' + n^2\psi + 8\pi
a^{-2}\psi\left[-\frac{1}{6}(\overline{\varepsilon} -
3\overline{p}) + 2\overline{p} + 2\overline{\tau} \right] = 0,
\end{equation}
where for convenience we have proceed to conserving on the
ultrarelativistic stage conform densities of energy and pressure
$\overline{\varepsilon} = a^4(\eta)\varepsilon$, $\overline{p} =
a(\eta)^4p$ etc.

For calculation of $\overline{\tau}(\eta)$ we will advert to
relativistic kinetic equations
\begin{equation}\label{1.9}
[H,f]  =J[f].
\end{equation}

The appearance of GW leads to the distortion of mass surface,
thereof it is necessary to produce a re\-nor\-ma\-li\-za\-ti\-on
of momentum \cite{Yukin2_0}-\cite{Yukin2}. It is simpler to make
an indicated re\-nor\-ma\-li\-za\-ti\-on by transformation
(\ref{1.9}) of momentum variables $p_i$ to the momentum variables
$\mathbb{P}_i$
\begin{equation}\label{1.10}
\mathbb{P}_i = p_i - \frac{1}{2}h^{k}_ip_k,\, p_i = \mathbb{P}_{i}
+ \frac{1}{2}h^k_i\mathbb{P}_k + O(h^2).
\end{equation}
The Jacobian of this transformation in case of traceless
perturbations of metric in linear by $h$ approach is equal to one.
In new variables the Hamilton function coincide with the
unperturbed value accurate to members $O(h^2)$:
\begin{equation}\label{1.11}
H(x,p) = \frac{1}{2}g^{ik}p_ip_k =
\frac{1}{2}\stackrel{o}{g^{ik}}\mathbb{P}_i\mathbb{P}_k =
\stackrel{o}{H}(x,p).
\end{equation}
Therefore $g = \stackrel{o}g + O(h^2),$ differential of volume of
mo\-men\-tum space is invariant with respect to transformation
(\ref{1.10})
$$
d\pi' = \frac{2s + 1}{(2\pi)^3}\,\frac{\alpha^4p}{\sqrt{-g}}
\delta(H - \frac{1}{2}m^2) = $$
\begin{equation}\label{1.12}
 = \frac{2s + 1}{(2\pi)^3}\, \frac{\alpha^4\mathbb{P}}{\sqrt{-g}}
\delta(\stackrel{o}{H} - \frac{1}{2}m) = d \stackrel{o}{\pi}.
\end{equation}
Taking into account the fact, that momentum variables can be
contained in invariant scattering amplitude $M_{if}$ by means of
all possible contractions of type $(p,p')$, it can be strictly
shown, that collision integral accurate to $O(h^2)$  is invariant
with respect to transformations (\ref{1.10}). Thus, representing a
distribution function $f(x,p)$ in form
\begin{equation}\label{1.13}
f(x^i,p_i) = f_0(\eta ,\mathbb{P}_4) + \delta f(x^i,\mathbb{P}_k),
\end{equation}
where $f_0(\eta ,\mathbb{P}_4)$ - isotropic solution of kinetic
equations (\ref{1.9}), and integrating the kinetic equations
(\ref{1.9}), we will obtain linearized kinetic equations for
distributions deviations $\delta f$:
\begin{equation}\label{1.14}
\mathbb{P}^i \frac{\partial \delta f}{\partial  x^i}
+\frac{1}{2}\frac{\partial f_0}{\partial
\mathbb{P}_0}\mathbb{P}_{\beta}\mathbb{P}^{\alpha}h'^{\beta}_{\alpha}
= J^1[f_0,\delta f].
\end{equation}
The isotropic part of distribution function $f_0$ satisfies to
zero approximation of equation (\ref{1.9}):
\begin{equation}\label{1.15}
\mathbb{P}^0 \frac{\partial  f_0}{\partial  \eta}
+\frac{a'}{a}(\mathbb{P}_0\mathbb{P})\frac{\partial f_0}{\partial
\mathbb{P}_0} = J^0[f_0].
\end{equation}
Calculating the MET perturbation with the account of
transformation (\ref{1.10}), we'll find:
\begin{equation}\label{1.17}
\delta T_{ik} = -ph_{ik} + \sum \int
\mathbb{P}_i\mathbb{P}_k\delta fd\pi,
\end{equation}
where summation is carried out by all kinds of particles,
participating in reactions. From linear equations (\ref{1.14}) is
clear, that $\delta f\sim S_{\alpha
\beta}\mathbb{P}^{\alpha}\mathbb{P}^{\beta}$. Actually, supposing
such dependence we arrive at con\-clu\-si\-on that collision
integral $J^1$ in consequence of it's invariance can be only a
linear combination of contractions of type $S_{\alpha
\beta}\mathbb{P}^{\alpha}\mathbb{P}^{\beta}$, $S_{\alpha
\beta}n^{\alpha}\mathbb{P}^{\beta}$, $S_{\alpha
\beta}n^{\alpha}n^{\beta}$, $S_{\alpha \beta}\delta^{\alpha
\beta}$, from which only the first is different from zero. And
integrals of type (\ref{1.17}) are defined by the only selected
spatial direction, $n_\alpha$. Therefore these integrals become
the linear combination of contractions:
$$S_{\gamma\delta}n^\gamma n^\delta,\; S_{\gamma\delta}\delta^{(\gamma\alpha}n^\beta n^{\gamma)},\;
S_{\gamma\delta}\delta^{(\alpha\gamma}\delta^{\beta\gamma)},$$
from which only the last is different from zero and equal to
$S_{\alpha\beta}$.

Thus, finally we'll obtain:
\begin{equation}\label{1.18}
\delta \overline{T}_{ik} = -\overline{h}_{ik}p +
\overline{h}_{ik}\overline{\tau}_f,
\end{equation}
where
$$\overline{\tau}_f = \frac{1}{S^2}\sum \int d\pi
\mathbb{P}_{\alpha}\mathbb{P}_{\beta}S_{\alpha \beta}\delta f$$
and the designation  $S^2 = S_{\alpha \beta}S_{\alpha \beta}$ is
incorporated, - summation is carried out by repeating indexes.

Thus, equation (\ref{1.8}) takes form:
\begin{equation}\label{1.19}
\psi'' + \eta^2\psi + 8\pi a^{-2}\psi
\left[-\frac{1}{6}(\overline{\varepsilon} - 3\overline{p}) +
2\overline{\tau}_f\right] = 0.
\end{equation}

\section{Calculation of collision integral}
Let's turn to calculation of $\overline{\tau}_f$. In case of
equilibrium distributions $f_0$ an integral $J^1$ can be
represented in form \cite{Yukin9}:
\begin{equation}\label{2.1}
J^1[f_0\delta f] = \frac{\delta f}{1\pm f_0}\mathbf{K}
(\mathbb{P},\eta),
\end{equation}
where
$$\mathbf{K}(\mathbb{P},\eta) = $$
\begin{equation}\label{2.3}
= \sum \int \prod\limits_{f,i}'d\pi \delta(\mathbb{P}_f -
\mathbb{P}_i)W_{if}\prod\limits_{f}'f_0\prod\limits_{f}'(1\pm
f_0). f_0)'
\end{equation}
Here summation is carrying out by all initial,(i),and final, (f),
states of given sort particles, participating in reactions;
$\prod_{f,i}'d\pi$ means the product of momentum volumes of all
particles, except given, $W_{if}$ - invariant scattering matrix
(details see in Ref. \cite{Yukin11}).

Let's illustrate a statement (\ref{2.1}) for four-particle
reactions $ab \longleftrightarrow cd$. In that case collision
integral takes form:
$$J_{ab\longleftrightarrow cd} =
\frac{1}{2(2S_a + 1)(2S_b + 1)}\times$$
$$\times \int \frac{d\pi_b}{2}\frac{d\pi_c}{2}\frac{d\pi_d}{2}(2\pi)^4\delta(\mathbb{P}_a
+ \mathbb{P}_b -  \mathbb{P}_c -\mathbb{P}_d
)\overline{|M_{if}|}^2\times$$
\begin{equation}\label{2.3}\times\left\{f_cf_d(1\pm f_a)(1\pm
f_b)- f_af_b(1\pm f_c)(1\pm f_d)\right\} .
\end{equation}
Multipliers $(2S_a + 1)^{-1}$, $(2S_i + 1)^{-1}$ â (\ref{2.3})
correspond to the average by polarized states of particles
\cite{Yukin12}, and coefficients $1/2$ respond to the selected
normalization of colliding particles wave functions
\cite{Yukin13}. Substituting $f = f_0 + \delta f$, where $f_0$
--- isotropic equilibrium distribution function, and $\delta f\sim
S_{\alpha \beta}\mathbb{P}^{\alpha}\mathbb{P}^{\beta}$,  and
linearizing an integral (\ref{2.3}) by $\delta f$, it is easy to
make sure that addends in the figure brackets of type $\delta
f_cf^0_d(1\pm f^0_a)(1 \pm f^0_b)$ during integration by $d\pi$
turn to zero. Actually, directing an axis $\mathbb{P}^c$ along the
wave vector $n_{\alpha}$ we'll achieve, that $\delta f_c$ will be
proportional to expressions $(\mathbb{P}^2_{c2} -
\mathbb{P}^2_{c3})$ and $\mathbb{P}_{c2} \mathbb{P}_{c3}$. Since
multipliers at $\delta f_c$ are invariant relatively to
transformations $\mathbb{P}_2 \rightarrow \mathbb{P}_3$ and
$\mathbb{P}_3 \rightarrow \mathbb{P}_2$ (for all particles
simultaneously), and $\delta f_c$ at that change a sign, then
during integration by momentums in infinite limits, this added
turns to zero. Non-zero contribution in $J^1$ will give only
members, containing $\delta f_a$, since integration doesn't carry
out by momentums $\mathbb{P}_a$. Thus, discarding non-sufficient
for hot model statistical mul\-ti\-pli\-ers, we'll find for
ultrarelativistic particles:
\begin{equation}\label{2.4}
\mathbf{K}(\eta ,p) = \nu(p,\eta)p,
\end{equation}
where effective frequency of collisions $\nu(p,\eta)$ is equal:
$$\nu(p,\eta)=$$ \begin{equation}\label{2.5}=\frac{(2S_c + 1)(2S_d + 1)}{16\pi^2(2S_a +
1)p^2}\int\limits_{0}^{\infty}f_0(q)dq\int\limits_{0}^{4pq}s\sigma_{tot}(s)ds;
\end{equation}
$p$ - absolute value of physical momentum,
$$\sigma_{t_0t}(s) = \frac{1}{16\pi S}
\int\limits_{0}^{1}dx|\overline{M(s,x)}|^2$$ - total scattering
cross-section,
$$x = -t/s, \quad s = (p_a + p_b)^2,\quad t = (p_a - p_c)^2 $$
- kinematic invariants.

When all interacting particles are ultrarelativistic, kinetic
equations (\ref{1.14}) take form:
$$\frac{\partial \delta f}{\partial \eta} + a\nu \delta f -
\frac{\mathbb{P}_{\alpha}}{\mathbb{P}}\frac{\partial \delta
f}{\partial x^{\alpha}} =$$
\begin{equation}\label{2.6}
 = -\frac{1}{2}\frac{\partial f_0}{\partial
\mathbb{P}}\frac{S_{\alpha
\beta}\mathbb{P}_{\alpha}\mathbb{P}_{\beta}}{\mathbb{P}}\left(\frac{\psi}{a}\right)'e^{-i\eta_{\alpha}x^{\alpha}}.
\end{equation}
It should be pointed out, that here $\mathbb{P}\equiv
\mathbb{P}_0$ differs from physical momentum by multiplier
$(\mathbb{P} = a(\eta)p_{phys})$. It is obvious, that at $\nu
> 0$ the second member in equation, conditioned by interparticle collisions,
leads to the decrease of $\delta f$, i.e., to the relaxation
towards the equilibrium distribution. Let's write down the
solution of equation (\ref{2.6}), turning to zero at $S_{\alpha
\beta} = 0$ and owning the structure of collisionless equation at
$\eta_0 \rightarrow 0$:
\begin{equation}\label{2.7}
\delta f = -\frac{1}{2}\frac{\partial f_0}{\partial
\mathbb{P}}\frac{S_{\alpha
\beta}\mathbb{P}_{\alpha}\mathbb{P}_{\beta}}{\mathbb{P}}e^{-i\eta_{\alpha}x^{\alpha}
- i\pi_{\alpha}-\gamma(\eta) }\times
\end{equation}
$$
\times \lim_{\eta_0\to 0}
\left\{\left(\frac{\psi}{a}\right)e^{i\pi_{\alpha}n_{\alpha}\eta_0}
+
\int\limits_{\eta_0}^{\eta}(\psi/a)'e^{\gamma(\eta')+i\pi_{\alpha}n_{\alpha}\eta'}d\eta'\right\},$$
where unit vector $\pi_{\alpha} = \mathbb{P}_{\alpha}/\mathbb{P}$
and {\it damping decrement} are incorporated:
\begin{equation}\label{2.8}
\gamma(\eta ,\mathbb{P}) = \int\limits_{\eta_0}^{\eta}\nu(\eta
,\mathbb{P})a(\eta)d\eta = \int\limits_{t_0}^t\nu(t,\mathbb{P})dt.
\end{equation}
It is not too hard to calculate the MET perturbation, conditioned
by $\delta f$,
$$\delta T_{\alpha \beta} = \sum_{a}\int
\mathbb{P}_{\alpha}\mathbb{P}_{\beta}\delta fd\pi =$$
$$= -\frac{\pi S_{\alpha
\beta}}{4
a^2}e^{-in_{\alpha}x^{\alpha}}\sum_{a}\frac{2S_a+1}{(2\pi)^3}
\int\limits_0^{\infty}d\mathbb{P}\mathbb{P}^4\frac{\partial
f_0}{\partial \mathbb{P}}e^{-\gamma(\eta ,\mathbb{P})}\times$$
$$\times \lim_{\eta_0\to 0}\left\{\left(\frac{\psi}{a}\right)_{\eta_0}J[\eta(\eta -
\eta_0)] +\right.$$
\begin{equation}\label{2.9}
+ \left. \int\limits_{\eta_0}^{\eta}e^{\gamma(\eta'
,\mathbb{P})}\left(\frac{\psi}{a}\right)^1J[n(\eta -
\eta')]d\eta'\right\}.
\end{equation}
The summation in (\ref{2.9}) is carried out by sorts of particles
$a$, participating in reactions, and incorporated function:
\begin{equation}\label{2.10}
J(x) = \frac{8}{x^2}\left[\frac{\sin x}{x}\left(\frac{3}{x^2} -
1\right) - \frac{3\cos x}{x^2}\right],
\end{equation}
having the asymptotics
\begin{equation}\label{2.11}
\stackunder{x\to \infty}{J(x)}
 \simeq -\frac{8\sin x}{x^3}, \quad J(0) = \frac{16}{15}.
\end{equation}
Let's consider now the relationships, that are correct on the
ultrarelativistic stage of Universe evolution (see., for example,
\cite{Yukin14}):
\begin{equation}\label{2.12}
\begin{array}{l}%
a = a_1\eta,\\
\\
\overline{T}^4 =(aT)^4 = {\displaystyle \frac{45}{4\pi^3N}a^2_1},\\
\\
\mbox{ãäå $N$ - the statistical factor of particles number:}\\
\\
N =
\sum_{B}(2S + 1) + \frac{7}{8}\sum_{F}(2S + 1).$$
\end{array}
\end{equation}
Then we'll obtain:
$$16\pi \overline{a}^2\overline{\tau}_f = $$
$$\frac{45}{8\pi^4N\eta}\sum_{a}(2S + 1)\int\limits_{0}^{\infty}dzz^4
\frac{\partial f_0}{\partial z}e^{-\gamma(z ,\eta)}\times$$
$$\times \lim_{\eta_0\to0}\left\{\left(\frac{\psi}{\eta}\right)_{\eta_0}J[\eta(\eta
- \eta_0)] + \right.
$$
\begin{equation}\label{2.13}
\left. +
\int\limits_{\eta_0}^{\eta}e^{\gamma(z,\nu'_0)}\left(\frac{\psi}{\eta'}\right)^1
J[\eta(\eta - \eta')]d\eta'\right\},
\end{equation}
where $z = \mathbb{P}/\overline{T}\equiv p/T.$ \vskip 12pt

Thus, finally we have - {\it the evolution of cosmological GW in
isotropic ultrarelativistic gas $(\varepsilon = 3p)$ defines by
equation:
\begin{equation}\label{2.14}
\psi'' + \eta^2\psi + 16\pi a^{-2}\overline{\tau}_f\psi = 0,
\end{equation}
where $\overline{\tau}_f$ is described by expression
(\ref{2.13})}.

\section{Extreme cases}
Let's at first consider the case of fast relaxation
\begin{equation}\label{3.1}
\gamma \gg 1.
\end{equation}
Producing the asymptotical estimation of integral (\ref{2.13}) by
Fourier method, we'll reduce an equation (\ref{2.14}) to the form
\begin{equation}\label{3.2}
\psi'' + \eta^2\psi +
\frac{8}{5}\left(\frac{\psi}{\eta}\right)^1\frac{1}{a\nu_{eff}\eta}
= 0,
\end{equation}
where
\begin{equation}\label{3.3}
\frac{1}{\nu_{eff}} = -\frac{{\displaystyle\sum(2S +
1)\int\limits_{0}^{\infty}dz\ z^4\frac{\partial f_0}{\partial
z}\frac{1}{\nu(z,\eta)}}}{{\displaystyle 4\sum(2S +
1)\int\limits_{0}^{\infty}dz\ z^3f_0}}.
\end{equation}
In consequence of condition (\ref{3.1}) the last member in the
left part of equation (\ref{3.2}) is small in comparison with the
first two members - it responds to the weak damping of GW vacuum
oscillations:
$$
h^{\alpha}_{\beta} =
-\frac{1}{a}\exp\left(-\int\limits_{0}^{\eta_0}\frac{4d\eta}{5a\eta^2\nu_{eff}}\right)\times
$$
\begin{equation}\label{3.4}
\times\left\{S^{+}_{\alpha \beta}e^{-i(n\eta
-n_{\alpha}x^{\alpha})} + S^{-}_{\alpha \beta}e^{i(n\eta
-n_{\alpha}x^{\alpha})}\right\}.
\end{equation}
At $\nu_{eff}t\rightarrow \infty$ the damping of GW vanishes: in
the ultrarelativistic fluid GW propagate as well as in vacuum.
This fact was mentioned in the beginning of the article.

Let now $\gamma \ll 1$. In the collisionless approximation an
equation (\ref{2.13}) can be simplified:
$$16\pi \overline{a}^2\overline{\tau}_f = \frac{3}{2\eta}\lim_{\eta_0\to
0}\left\{\left(\frac{\psi}{\eta}\right)_{\eta_0}J[\eta(\eta -
\eta_0)] + \right.
$$
\begin{equation}\label{3.5}
\left. +
\int\limits_{\eta_0}^{\eta}\left(\frac{\psi}{\eta'}\right)^1J[\eta(\eta
-\eta')]d\eta'\right\}.
\end{equation}
This expression have two asymptotics. In the long-wavelength limit
$(n\eta \ll 1)$ we'll find
\begin{equation}\label{3.6}
16\pi a^{-2}\overline{\tau}_f\simeq
\frac{8}{5}\frac{\psi}{\eta^2}.
\end{equation}
Substituting (\ref{3.6}) in the equation {\ref{2.14}}, we'll
obtain re\-se\-ar\-c\-hed in \cite{Yukin3} oscillations, as a
solution:
$$\psi =\sqrt{\eta}\ \times$$
\begin{equation}\label{3.7}
\left[C_+\cos \left(\frac{3\sqrt{3}}{2}\ln n\eta \right) + C_-\sin
\left(\frac{3\sqrt{3}}{2}\ln n\eta \right) \right].
\end{equation}
Such behavior, however, is forming only in the case that
collisionless situation occurred from the very beginning at $\eta
= 0$. And if strong-collision phase pre\-ce\-ded the collisionless
phase (down to $\eta = \eta_0)$, then instead of (\ref{3.6}) we
have now:
\begin{equation}\label{3.8}
16\pi a^{-2}\overline{\tau}_f\simeq
\frac{8}{5}\left[\frac{\psi(\eta)}{\eta^2} -
\frac{\psi(\eta_0)}{\eta_0\eta}\right],
\end{equation}
and equation (\ref{2.14}) becomes inhomogeneous
\begin{equation}\label{3.9}
\psi'' + \frac{8}{5}\frac{\psi}{\eta^2} =
\frac{8}{5}\frac{\psi(\eta_0)}{\eta_0\eta}.
\end{equation}
It's solution, satisfying to the sewing condition at the moment
$\eta = \eta_0$, has a form:
\begin{equation}\label{3.10}
\psi(\eta) = \psi(\eta_0)\frac{\eta}{\eta_0} +
A\sqrt{\frac{\eta}{\eta_0}}\sin \left(\frac{3\sqrt{3}}{2}\ln
\frac{\eta}{\eta_0}\right).
\end{equation}
Main part of GW amplitude $h^{\alpha}_{\beta}$ at that remains
constant. Thus, a solution for long GW's, obtained in Ref.
\cite{Yukin3}, is true only in the case that Universe started from
the collisionless phase, and is false in the case of existing of
initial hydrodynamic stage. At the last case GW stores it's final
condition on hydrodynamic stage.

\section{The evolution of short waves}
Let's consider the evolution of short $(n\eta \gg 1)$ GW's at
arbitrary $\gamma(\eta ,\mathbb{P})$. In this case it is
convenient to lay:
\begin{equation}\label{4.1}
\psi = \widetilde{\psi}(\eta)^0\exp \left(-i \int \Omega d\eta
\right),
\end{equation}
where $\Omega(\eta)$ - is a large value, $\widetilde{\psi}(\eta)$
- slowly changing function. At $\eta \to 0$ GW's with any $\eta$
are long, therefore it is necessary to redefine the solution
(\ref{2.7}) at a point of time $\eta_0$, when GW's become short:
$\eta_0 \gtrsim \eta^{-1}$. Carrying out necessary calculations,
we'll find in the case of ultrarelativistic particles:
$$16\pi a^{-2}\overline{\tau}_f =
-\frac{3}{4}\frac{\Omega}{n^4\eta^2}\sum \frac{2s +
1}{2\pi^2}\int\limits_{0}^{\infty}dzz^4\frac{\partial
f_0}{\partial z}\Omega'\left\{\frac{2}{3}n^2 -\right.$$
\begin{equation}\label{4.2}
\left.- (\Omega'^2 - n^2) + \frac{\Omega'^2 - n^2}{2\Omega'n}\ln
\left|\frac{\Omega' + n}{\Omega' - n}\right|\right\},
\end{equation}
where $\Omega' = \Omega + i\nu(\eta ,p)a(\eta)$. For the
weak-collision plasma $\Omega\approx n$, holding in
Ref.(\ref{4.2}) members $\Omega^2 - n^2$, we'll reduce an equation
(\ref{2.14}) to the form:
\begin{equation}\label{4.3}
\widetilde{\psi}'' - 2i\Omega \widetilde{\psi}' - (\Omega^2 -
n^2)\widetilde{\psi} - 4\frac{ia\nu_{eff}}{n\eta^2}= 0,
\end{equation}
where $\nu_{eff}$ defines analogically by (\ref{3.3}). In
mentioned approach the solution of equation (\ref{3.4}) is:
\begin{equation}\label{4.4}
\begin{array}{l}
\Omega^2\simeq n^2 + \frac{2}{\eta^2},\\
\widetilde{\psi}\simeq k^{1/2}(\eta) =\exp
\left(-\frac{1}{2}\gamma_{G}(\eta)\right),\\
\end{array}
\end{equation}
where
\begin{equation}\label{4.5}
\gamma_G(\eta) =
\frac{4}{\eta^2}\int\limits_{\eta_0}^{\eta}\frac{a\nu_{eff}}{\eta^2}d\eta
= \int\limits_{t_0}^{t}\frac{\nu_{eff}dt}{k^2t^2}.
\end{equation}
Let's point an inquisitive fact: neglecting damping of GW and
introducing value $\Phi = \psi /a$, we'll obtain from (\ref{4.3})
an equation for the mass scalar field
$$\Delta_2\Phi - \frac{R}{6}\Phi + m^2_G\Phi = 0,$$
with the effective mass of graviton $$m_G = \frac{1}{\sqrt{2}t}.$$
Let's clarify the question about GW evolution, which make sense
only in the approximation of geometrical optics. Averaged
pseudotensor of momentum energy is equal to
\begin{equation}\label{4.6}
< t^{ik} > = \frac{1}{32\pi} < h^{n,i}_q\overline{h}\ ^{q,k}_n >.
\end{equation}
Substituting here (\ref{1.6}), we'll obtain MET of perfect
ultrarelativistic fluid with energy density
\begin{equation}\label{4.7}
\varepsilon_G = \frac{s^2n^2}{a^4}\exp[-\gamma_G(\eta)],
\end{equation}
i.e., the value $\gamma_G$ (or $k(\eta)$) can be named as a
decrement of GW energy damping.

\section{An example of GW energy damping decrement calculation
in the reactions of type $ee^+\rightarrow$ hadrons} For
illustration we'll carry out a calculation of GW damping according
to stated above scheme for reactions of type $ee^+\to$ hadrons,
the cross-sections of which have a scaling behavior \cite{Yukin15}
\begin{equation}\label{5.1}
\sigma_{t_0t}(s) = \frac{4\pi \alpha^2}{5s}\sum e^2_i\equiv
\frac{4\pi \alpha^2Q^2(s)}{3s},
\end{equation}
where $e_i$- fundamental charges; their number, and, therefore the
value $Q$, weakly depend from $s$. After calculations we'll find
\begin{equation}\label{5.2}
\nu_{eff}(s) = \frac{45\zeta(3)N\alpha^2}{2\pi^3}TQ^2(T),
\end{equation}
\begin{equation}\label{5.3}
\gamma_{G}(\infty) = \frac{4(a\nu_{eff})\eta_0}{n^2\eta_0}\sim
\frac{1}{\eta_0}.
\end{equation}
As it was mentioned above, the GW damping is possible during
synchronous fulfillment of conditions $n\eta
> 1$ (short GW's) and $e_{eff}a < n$ (weak-collision gas). In the opposite case
we have either long GW's, or GW's in the perfect liquid; in the
both cases damping is absent. Therefore, regarding the fulfillment
of these conditions in the initial point of time $\eta_0$, we
arrive to the conclusion, that $\gamma_G(\infty)\simeq 4\beta^2$,
where $\beta < 1$ - unknown factor, guaranteing the severity of
made approximations, moreover all GW's, having the wavelength
\begin{equation}\label{5.4}
\lambda \gtrsim {\displaystyle \frac{\lambda_\gamma \beta
2\pi^3}{45\zeta(3)Q^2(t_0)N\alpha^2}} \sim 10^4\lambda_\gamma,
\end{equation}
on the actual point of time, damp with the same decrement (where
$\lambda_\gamma$ - wavelength of relict photons).

\section{The results of GW damping decrement calculation in the different reactions}
Let's diagram the evolution of GW in hot Universe as a dependence
from the behavior of total cross-section of scattering. Let's
isolate following typical situations:

\begin{itemize}
\item[A:] $n\eta \ll 1, \quad a\nu \eta \gg 1$ - long waves in the
liquid,
\item[B:] $n\eta \gg 1, \quad a\nu \eta \gg 1$ - short waves in the
liquid,
\item[C:] $n\eta \ll 1, \quad a\nu \eta < 1$ - long waves in the
weak-collision gas,
\item[D:] $n\eta \gg 1, \quad a\nu \eta < 1$ - short waves in the
weak-collision gas.
\end{itemize}

$$\begin{array}{l}
\sigma_{t_0t}(s)\sim \frac{\alpha^2}{s}\\
n_0\sim \alpha^2\\
\end{array}
\begin{array}{llll}
 &n>n_0 &C\rightarrow \Osymbol{D} &\\
\mbox{{\Large$\nearrow$}} & & &\\
\mbox{{\Large$\searrow$}} & & &\\
 &n<n_0 &C\rightarrow A & \rightarrow B\\
\end{array}$$
$$\sigma_{t_0t}\sim \frac{\alpha^2}{m\sqrt{s}}
\begin{array}{llll}
 &\alpha^2<m &C\rightarrow\qquad \Osymbol{D} &\\
\mbox{{\Large$\nearrow$}} & & &\\
\mbox{{\Large$\searrow$}} & & &\\
 &\alpha^2>m &A\longrightarrow \quad B & \quad\longrightarrow \Osymbol{D}\\
\end{array}
$$
$$\begin{array}{r}
\sigma_{t_0t}\sim
\frac{\alpha^2}{m^2}\left(\frac{s}{m^2}\right)^{\gamma}\\
n_0\sim
\frac{m^2}{\alpha^2}\left(\frac{m}{\alpha^2}\right)^{\frac{2}{2\gamma +1}}%
\end{array}
\begin{array}{llll}
 &n>n_0 &A\rightarrow B& \rightarrow \Osymbol{D}\\
\mbox{{\Large$\nearrow$}} & & &\\
\mbox{{\Large$\searrow$}} & & &\\
 &n<n_0 &A\rightarrow \fbox{C} & \rightarrow \Osymbol{D}\\
\end{array}
$$
Situations, where collisional GW damping occur are circled; the
domain of GW amplitude fixation are squared. The problem of
damping effectiveness requires the realization of calculations for
concrete reactions.

The results of such calculations, fulfilled in papers
\cite{Yukin7}-\cite{Yukin9}, are represented in the Table 1. In
two last columns of the Table represented the results, summarized
by all reactions of given type. $G$ - Fermi constant,
$$\sigma = \frac{0,1\alpha_XN^2m_{p1}}{m_X},$$ $m_{p1}$ -
Planck mass, $m_X$ - mass of $X$ - bosons, $\alpha_X$ -
interaction constant, $N_X$ - the number of particles
participating in reactions.

Summing the results, we'll note, that collision dam\-ping of
cosmological GW's in the all conceivable re\-ac\-ti\-ons is not
great. However, the reason of weakness of GW damping isn't the
weakness of interparticle interaction, but the circumstance, that
exactly in the medium with intense interparticle interactions,
i.e. in the perfect fluid, GW's don't damp. In order to provide an
appreciable GW damping, collisions don't need to be very frequent,
that, in turn, leads to the weakness of damping. \pagebreak
\begin{center}
{\bf Table 1.} The damping decrement of cosmological gravitational
waves\\ in the early hot universe for different processes in
cosmological plasma \vskip 12pt

\begin{tabular}{|l|l|l|l|}
\hline
\parbox{2.5cm}{Process}&
\Boxf{3}{Total cross-section of scattering\\ {\normalsize
\centerline{$\sigma_{t_0t}(s)$}}}& \Boxf{3}{The region of
modern wavelengths, where damping is maximal\\
(in cm)} &
\Boxf{3}{Ma\-xi\-mal value $\gamma_G$}\\
\hline
\Boxn{2.5}{$\begin{array}{l}%
e\gamma^+ \leftrightarrow e\gamma\;^{1)}\\
ee^+\leftrightarrow \gamma \gamma\\
\end{array}$}&
\Boxn{3}{${\displaystyle \frac{2\pi\alpha^2}{s}\left(\ln
\frac{s}{m^2} + \frac{1}{2}\right)}$}&\Boxn{3}{\centerline{$\sim
100$}}
&\Boxn{2}{ \centerline{$0,01$}}\\
\hline
\Boxn{2.5}{$\begin{array}{l}%
ee^+ \leftrightarrow \mu \mu^+\;^{2)}\\
ee^+\leftrightarrow \mbox{àäðîíû}\\
\end{array}$}&
\Boxn{3}{${\displaystyle
\frac{4\pi\alpha^2}{3s}Q^2(s)}$}&\Boxn{3}{${\displaystyle\lambda
> \frac{10^4\beta^2\lambda_{\gamma}}{Q^2N'}\sim 10^3}$}
&\Boxn{2}{ \centerline{$4\beta^2$}}\\
\hline
\Boxn{2.5}{$\begin{array}{l}%
e\nu \leftrightarrow e \nu \;^{3)}\\
\\
\\
ee^+\leftrightarrow \nu\overline{\nu}\\
\\
\\
e\overline{\nu}\leftrightarrow e\overline{\nu}\\
\\
\nu_{\mu}e\leftrightarrow \nu_e\mu\\
\\
\nu_{\mu}\overline{\nu}_e\leftrightarrow e_+\mu^-\\
\\
\nu_{\mu}\mu_+\leftrightarrow e_+\nu_e\\
\end{array}$}&
\Boxn{3}{
$\begin{array}{l}%
{\displaystyle G^2_Fs(g^2_L +
\frac{1}{3}g^2_R)/\pi}\\
\\
{\displaystyle G^2_Fs(\frac{1}{3}g^2_L +
\frac{1}{3}g^2_R)/\pi}\\
\\
{\displaystyle G^2_Fs(\frac{1}{3}g^2_L +
g^2_R)/\pi}\\
\\
G^2_Fs/\pi\\
\\
G^2_Fs/3\pi\\
\\
G^2_Fs/3\pi\\
\end{array}$}&\Boxn{3}{\centerline{$\sim
2\cdot 10^{20}$}}
&\Boxn{2}{ \centerline{$0,74$}}\\
\hline
\Boxn{2.5}{$\begin{array}{l}%
X\leftrightarrow \overline{q}\,\overline{q}\;^{4)}\\
X\leftrightarrow ql \\
\end{array}$}&
\Boxn{3}{$\begin{array}{c}%
|M|^2 = 8\pi
 \alpha_xm^2_xN\\
\sigma >1\\
\end{array}$
 }&\Boxn{3}{\centerline{${\displaystyle\frac{\sigma^{1/3}}{\alpha_XN}}$}}
&\Boxn{2}{ ${\displaystyle 0,29\frac{N_X}{N}\sim 3\cdot 10^{-2}}$}\\
\hline
\end{tabular}
 \vskip 12pt $^{1)}$By data of Ref.\cite{Yukin10};
 $^{2)}$by data of Ref.\cite{Yukin7},\cite{Yukin8};
 $^{3)}$by data of Ref. \cite{Yukin10};
 $^{4)}$by data of Ref. \cite{Yukin10},\cite{Yukin16}.
\end{center}
\vskip 12pt

The unique sufficiently effective (in order of GW damping)
interactions are the interactions with scaling behavior of
cross-section. Exactly scale-invariant interactions in consequence
of identical time law of collision frequency with GW frequency can
infinitely long influence on GW and thereby apply to the
sufficient damping of GW. However at attempt to calculate a GW
damping decrement in the most effective region $\nu \sim k$ we'll
fall into long-length collisionless phase, in which the GW energy
determination is difficult to define unambiguously. The
calculations of damping decrement during electro-weak interactions
generally confirmed the Hawking's estimation \cite{Yukin5} and
rather improve it. Realized calculations point that the height of
energetic spectrum of cosmological GW's can be decreased
approximately in 1,5 - 2 times at lengths $\lambda \gtrsim
10^4\beta^2/N'Q^2$cm and additional to this the same effect at
$\lambda\sim 2\times 10^{20}$cm.

\end{document}